\documentclass[%
 reprint,
 amsmath,amssymb,
 aps,
]{revtex4-2}

\usepackage{graphicx}% Include figure files
\usepackage{dcolumn}% Align table columns on decimal point
\usepackage{bm}% bold math
\usepackage{color}

\usepackage[colorlinks=true,linkcolor=magenta,citecolor=cyan]{hyperref}

\begin{document}

\preprint{APS/123-QED}

\title{Molecular Dynamics Investigation of Static and Dynamic Interfacial Properties in Ice–Polymer Premelting Layers}

\author{Takumi Sato}
%\email{sato8322@keio.jp}
\author{Ikki Yasuda}%
%\email{ikki8638@keio.jp}
\author{Noriyoshi Arai}%
%\email{arai@mech.keio.ac.jp}
\author{Kenji Yasuoka}%
\email{yasuoka@mech.keio.ac.jp}
\affiliation{%
 Department of Mechanical Engineering, Keio University, Yokohama, Kanagawa, 223-8522, Japan.\\
}

\date{\today}

\begin{abstract}
Premelting at the ice–polymer interfaces, in which a quasi-liquid layer (QLL) forms below the melting point, is strongly influenced by polymer surface chemistry; however, the molecular-scale mechanisms underlying these effects remain poorly understood. This study employs large-scale molecular dynamics simulations combined with machine learning-assisted analysis to elucidate how polymer type (hydrophilic vs hydrophobic) modulates interfacial premelting. Our simulations reveal that hydrophilic and hydrophobic polymer surfaces have distinct effects on the QLL thickness, interfacial water structure, and diffusivity. Specifically, a hydrophilic polymer interface promotes a thicker QLL with more ordered interfacial water and lower diffusivity, whereas a hydrophobic interface induces a thinner QLL with a less ordered interfacial water structure and higher diffusivity. These results advance the understanding of polymer-mediated interfacial melting phenomena and offer guidance for designing anti-icing and low-friction materials.
\end{abstract}

                              
\maketitle

\section{Introduction}
A quasi-liquid layer (QLL), characterized by structural disorder and enhanced molecular mobility, forms at the surface of crystalline ice as the temperature approaches the melting point from below. This phenomenon, known as premelting, is intrinsic to ice and has been implicated in a wide range of macroscopic processes including glacier dynamics, frost heave, trace gas adsorption, and low-friction behavior on icy surfaces\cite{michaelides2017,slater2019,nagata2019}. Although the QLL was identified over a century ago, its microscopic nature – structure, dynamics, and sensitivity to interfaces – remains an active area of investigation. Experimental studies employing spectroscopy and X-ray scattering \cite{maruyama2000,sazaki2010elementary,lin2023}, along with molecular simulations \cite{conde2008,limmer2014,kling2018,cui2022,zeng2024}, have revealed complex interfacial phenomena such as bilayer melting \cite{sanchez2017}, structural and dynamical heterogeneity \cite{pickering2018grand,yasuda2024}, and anisotropic growth across crystal \cite{llombart2020surface}. However, most such studies are limited to clean ice surfaces or idealized interfaces, leaving open questions about how the QLL behaves in more realistic, chemically-heterogeneous environments.

Theoretical treatments of premelting have emphasized the role of hydrogen bond frustration, entropic contributions, and long-wavelength surface fluctuations in determining the extent and thermodynamic stability of the QLL \cite{limmer2014,qiu2018,cui2023a}. While these mechanisms have been validated for bare ice–vapor or ice–solid interfaces \cite{shepherd2012,baran2022ice}, considerably less is known about how polymeric materials—ubiquitous in technological and environmental settings—modify the structure and dynamics of the QLL. Polymer–ice interfaces introduce additional degrees of freedom through their chain conformations, interfacial tension, and chemical functionality, all of which may influence interfacial water structure. Moreover, water–polymer interactions are known to be highly sensitive to the hydrophilic or hydrophobic character of the polymer, suggesting a strong dependence of QLL behavior on polymer chemistry. Some experimental works reported polymer-induced effect such as the anti-freezing coating \cite{chen2017icephobic,tamamoto2025anti} and enhanced ice premelting by hydrophobic polymers \cite{pallbo2024enhanced}. Previous molecular simulation studies involving small solutes have shown that ions and organic molecules can induce local restructuring and mobility gradients in the premelting layer \cite{hudait2017,niblett2021}, but the molecular effects of high-molecular-weight polymers remain largely unexplored.

The effects of polymer chemistry and surface topography on ice interfaces have been studied in recent works. For instance, amphiphilic polymer coatings, having both hydrophilic and hydrophobic domains, can effectively reduce ice adhesion by forming a quasi-liquid layer (QLL) at the interface \cite{mossayebi2023reduced}. Skountzos et al. showed that cross-linked polymer substrates, such as epoxy surfaces, induced structural disorder at ice interfaces due to hydrogen bonding with hydroxyl groups using molecular dynamics simulations  \cite{skountzos2024interfacial}. Surface roughness additionally influences ice adhesion. Fine surface textures can decrease ice adhesion by reducing the effective contact area, whereas excessive roughness can have the opposite effect, enhancing adhesion through mechanical anchoring of ice crystals \cite{meuler2010relationships}. These studies suggest that polymer chemistry and nanoscale surface structures of polymers influence the premelting layers \cite{zheng2022ice}, but the detailed molecular mechanisms is largely unexplored.

In this study, we investigated how polymer chemistry modulates the structural and dynamic properties of premelting layers at ice interfaces by means of large-scale, all-atom molecular dynamics simulations. Three representative polymers were selected to probe the influence of interfacial hydrophilicity: poly(ethylene oxide) (PEO) and poly(vinyl alcohol) (PVA), both hydrophilic, and polystyrene (PS), a prototypical hydrophobic polymer. Simulations were conducted over a range of temperatures spanning the melting point of ice to approximately 20 K below it, enabling direct comparison of QLL development under near-melting and subcooled conditions. Particular emphasis was placed on evaluating the diffusivity of water molecules in the interfacial region, a key dynamic metric that reflects the extent of surface melting and molecular mobility. Using machine learning-assisted analysis, we identified polymer-specific effects on premelting layer thickness, interfacial water structuring, and depth-dependent diffusivity. The results offer molecular-level insights into how polymer–ice interactions govern QLL behavior and could inform the design of materials for ice adhesion, lubrication, and thermal regulation applications.

\section{Method and Models}
\subsection{Modeling ice and polymers}
Using the open-source software GenIce \cite{Matsumoto2017}, a proton-ordered ice crystal was generated. This crystal consists of 27,648 water molecules, with approximate dimensions of 9.4 nm \(\times\) 8.9 nm \(\times\) 10.9 nm. The TIP4P/ICE \cite{abascal2005potential} model was employed to represent the water molecule, with an initial ice structure generated for each density at each temperature. The $x$-, $y$-, and $z$-axes are perpendicular to the primary, basal, and secondary prism planes, respectively. The ice-polymer interface was created by extending the unit cell in one direction from the initial structure and placing the polymer in that direction. The polymer was annealed by repeating the simulation twice, heating up from 259 K to 1000 K in 5 ns and cooling down to 259 K in 5 ns. 
All polymer models were simulated using the second-generation general AMBER force field (GAFF2)\cite{Wang2004}, which has a proven to be useful as a study of the mechanical properties at the interface of epoxy resins\cite{Konrad2023} and the frictional properties of fatty acid esters and steel surfaces\cite{Pominov2021}.

\subsection{Molecular dynamics simulation}
%=== copy & paste from kojima paper 
Molecular dynamics simulations were performed in the $NVT$ ensemble at $T$ = 254, 259, 264~K with a target normal pressure $p_z$ = 0.1 MPa. Instead of directly controlling $p_z$ via pressure coupling, we applied a constraint force between atoms in the central region of the ice and atoms in the uppermost region of polymers. The temperature range was chosen to be below the melting point of the TIP4P/Ice model (269.8 $\pm$ 0.1~K)\cite{Conde2017}. The integration time step was 1~fs, and after equilibration runs of more than 50~ns, production simulations were conducted for an additional 50~ns. Electrostatic interactions were evaluated using the particle mesh ewald method\cite{essmann1995smooth} with a short-range cutoff of 1.13 nm. Short-range van der Waals interactions were cut off at 1.13 nm. Bond lengths involving hydrogen atoms were constrained using the LINCS algorithm\cite{hess1997lincs}. Temperature was controlled using the stochastic velocity rescaling method\cite{bussi2007canonical} with a coupling time constant of 0.1~ps. All simulations were performed with GROMACS 2022.4\cite{abraham2015gromacs}, and trajectories were visualized using VMD \cite{humphrey1996vmd}.

%\subsection{Density profile} %もし必要であれば
%\subsection{Water permutation into polymer} %もし必要であれば 
%\subsection{Mean-square displacements (MSD)} %もし必要であれば

\subsection{Analysis of short-term dynamics using machine learning}
Previously, Yasuda et al.~\cite{yasuda2024} developed a machine-learning method to evaluate the short-term dynamics of water molecules in ice premelting layers. In this work, we employ this method to polymer-inducing effect on the premelting molecules at single-molecule resolution. In the machine learning approach, the short-term trajectory of oxygen atoms in the water molecules, $\bm{x}$, is input into a neural network, which outputs a score, $g(\bm{x})$, that quantifies the difference from the solid bulk water system, \begin{equation} 
    g(\bm{x}) = \mathbb{E}_{\bm{x}' \sim \bm{y}'} \left[ f^*(\bm{x}) - f^*(\bm{x}') \right] 
\end{equation} 
Here, $\mathbb{E}$ represents the expectation value taken over samples $\bm{x}'$ sampled from the distribution of $\bm{y}$, and $f^*(\bm{x}')$ is the function represented by the neural network, which is a simple multi-layer perceptron. For simplicity, $g(\bm{x})$ is a scalar value that is non-linearly converted from the multi-dimensarai
ional time-series data of a single oxygen atom position. The neural network model trained in ref.~\cite{yasuda2024} was used. Higher values of $g(\bm{x})$ correspond to molecular dynamics that differ from the solid phase.

\section{Results and Discussion}

\subsection{Interfacial Ice Structure and Premelt Layer Thickness}

Initially, we investigated the density distribution at the ice–polymer interface to examine the polymer type's impact on the melting behavior of ice. For clarity in comparing the density profile changes, density profiles were calculated using only the positions of oxygen atoms. Fig.~2 illustrates the density distributions of ice and polymer for each polymer at 254~K. Clear, well-defined peaks are visible in the bulk ice region ($L<30$). Here, $L$ is the position in the $y$-direction. However, approaching the ice–polymer interface, the density peaks gradually diminish, indicative of melting even below the melting point due to interactions with the polymer. At the interface, melting is particularly pronounced, as evidenced by the altered peak shapes. The first peak heights at 254~K decrease in the order PVA~$>$~PEO~$>$~PS, indicating the highest degree of melting at the PS interface.

Fig.~3 illustrates density distributions at 259~K, revealing further peak reductions compared to 254~K, reflecting increased melting at higher temperatures. Notably, the first peak heights at this temperature reorder to PVA~$>$~PS~$>$~PEO. This change demonstrates more significant melting at the PEO interface at 259~K compared to 254~K. The polymer density profiles near the interface show that PEO has penetrated significantly into the first ice layer, explaining the pronounced melting observed at this polymer–ice interface.

Density profiles at 264~K (Fig.~4) indicate further melting at the polymer–ice interface, particularly for PEO, where polymer chains penetrate as profoundly as the third ice layer. This suggests a pronounced influence of polymer penetration on interfacial melting. We quantified water ingress into each polymer phase to further elucidate the effects of polymer penetration.

\subsection{Influence of Polymer Properties on Water Ingress}

Fig.~5 presents the temporal evolution of the number of water molecules penetrating each polymer. The hydrophilic polymers PVA and PEO exhibited progressively increased water penetration over the simulation period, while no water penetration was observed for the hydrophobic polymer (PS). The absence of water penetration into PS is due to unfavorable hydrophobic interactions, effectively preventing water molecules from diffusing into the polymer's internal regions.

Interestingly, despite both being hydrophilic, significantly more water molecules penetrated PEO than PVA. To clarify the origins of this difference, polymer mobility was analyzed through mean-square displacement (MSD) calculations, shown in Fig.~6. At the simulation temperature of 259~K, only PEO is above its glass transition temperature ($T_g$), resulting in considerably higher mobility than the glassy PS and PVA. Thus, the diffusion coefficient of PEO is notably higher, underscoring the critical role polymer mobility and glass transition temperature play in facilitating water penetration and subsequent premelt layer formation.

\subsection{Classification of Interfacial Water Behavior Using Machine Learning}

To gain deeper insights into premelt layer formation dynamics at the ice–polymer interface, we applied machine-learning techniques to classify each ice layer as either solid-like or liquid-like based on the dynamical behavior of water molecules\cite{yasuda2024}. Fig.~7 shows classification results at 259~K for the ice–vapor interface and the three ice–polymer interfaces.

At the ice–vapor interface [Fig.~7(a)], Layer~1 exhibits liquid-like behavior, whereas Layers~3 and~4 are clearly solid-like. Layer~2 displays intermediate characteristics corresponding to the quasi-liquid premelt layer (QLL). Compared to the ice–vapor interface, water molecules at all polymer interfaces exhibited relatively more solid-like behavior, likely due to spatial confinement effects exerted by polymer chains.

For the hydrophilic polymer PVA [Fig.~7(b)], all layers near the interface were predominantly classified as solid-like. A similar observation was made for the other hydrophilic polymer, PEO [Fig.~7(c)], where Layers~1 and~2 also showed predominantly solid-like characteristics. This solid-like behavior is attributed to hydrogen bonding interactions between hydrophilic polymer chains and interfacial water molecules, significantly restricting water mobility and promoting structural ordering.

Conversely, in the case of the hydrophobic polymer (PS) [Fig.~7(d)], Layers~1 and~2 demonstrated clear premelt-like, liquid-like behavior. The hydrophobic polymer restricts water molecules solely by spatial confinement without additional hydrogen bonding constraints. This allows greater water molecule mobility, thereby facilitating the formation of an extended premelted layer at the polymer–ice interface.

These results collectively highlight the contrasting roles of polymer hydrophilicity and hydrophobicity in governing water molecules' dynamic and structural properties at ice interfaces. Hydrophobic polymers primarily induce premelt formation through spatial constraints, whereas hydrophilic polymers further restrict molecular mobility via hydrogen bonding, maintaining solid-like structures. Similar trend has been observed in a recent atomistic molecular simulations for polymer--ice interface \cite{skountzos2024interfacial}. Such detailed molecular-level insights enable improved understanding and rational design of polymer-modified surfaces tailored for specific ice adhesion, lubrication, and thermal management applications.

\section{Conclusion}
This study investigated the effects of polymer hydrophilicity and hydrophobicity on ice structure and water molecule behavior at the ice–polymer interfaces. The interfacial ice density peak was significantly lower at the interface with hydrophilic polymer PEO, indicating enhanced surface premelting. This effect was more pronounced at higher temperatures, suggesting that hydrophilic polymers facilitate premelt layer formation.

Additionally, some water penetrated into the hydrophilic polymer phase, with penetration extent depending on the polymer's glass transition temperature ($T_\mathrm{g}$). With a high $T_\mathrm{g}$, PVA exhibited limited polymer chain mobility, suppressing water infiltration.

Furthermore, we employed a machine-learning classification of solid-like vs. liquid-like water based on molecular dynamics, quantitatively evaluating spatial confinement effects at interfaces. At the hydrophobic polymer interface (PS), a thick premelt layer formed due solely to spatial confinement (no hydrogen bonding). In contrast, hydrogen bonding further restricted water mobility at hydrophilic polymer interfaces (PEO and PVA), maintaining solid-like ordering across several layers.

These findings deepen our understanding of how polymer properties influence ice–interface structure and ice melting and freezing processes. They also offer valuable insights for designing anti-icing surfaces and developing materials for low-temperature environments \cite{chen2017icephobic,zheng2022ice}. Understanding polymer-modulated premelting could help elucidate and control frictional properties on icy surfaces \cite{lecadre2020ice}.

\bibliography{ref}% Produces the bibliography via BibTeX.

\clearpage

%Table and Figures

%メモ: 図の埋め込みがされていないので，図の埋込をして(a), (b)を付記．(a)の図の色をあわせる．The melting temperature ($T_\mathrm{m}$) of the TIP4P/Ice model is 269.8 $\pm$ 0.1 K\cite{Conde2017}.
\begin{figure*}[htbp]
 \centering
\includegraphics[width=0.8\linewidth]{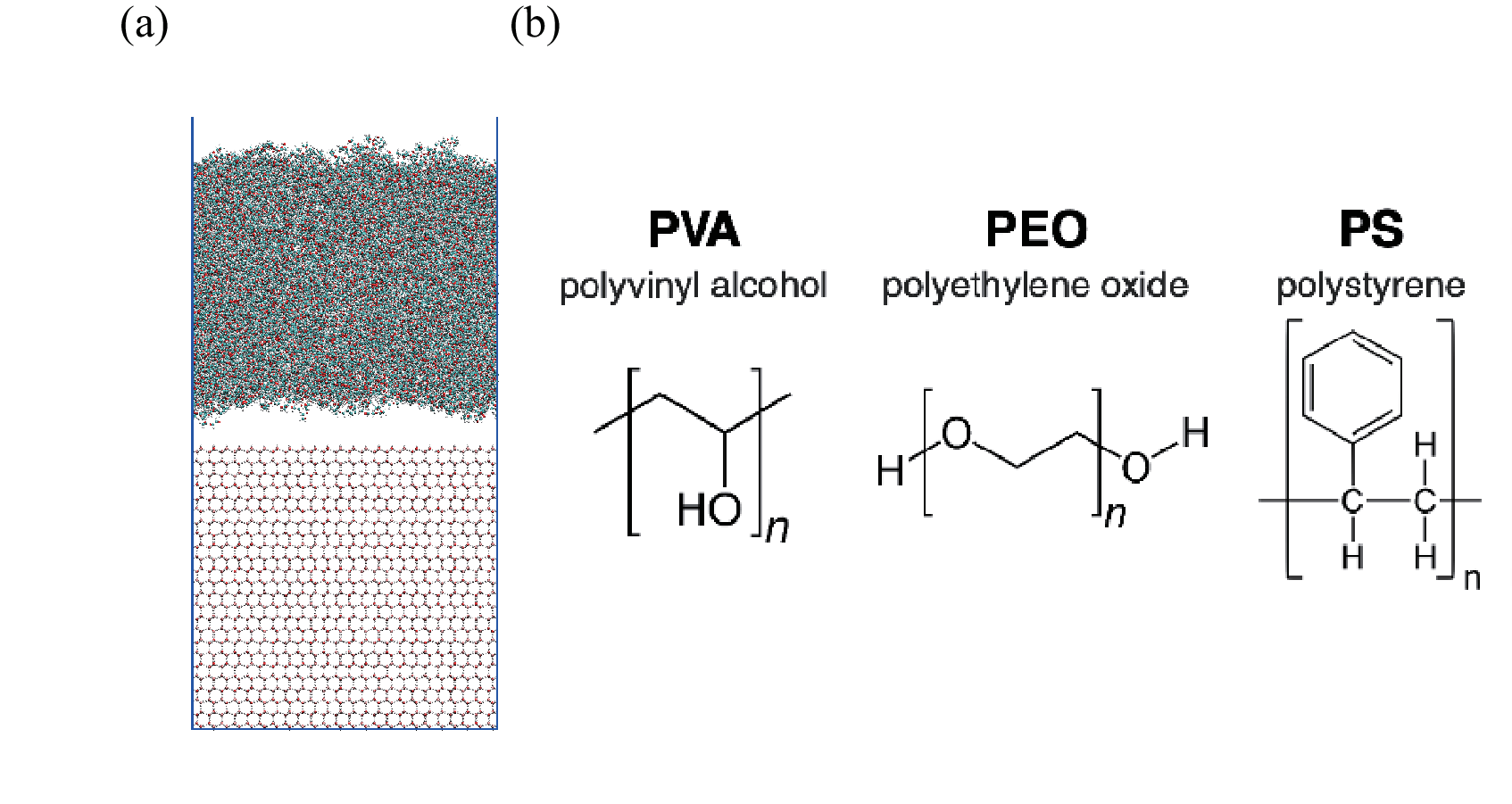}
\caption{(a) Initial configuration of the representative system. (b) Chemical structures of the polymer and water molecule model used in the simulation. }
\label{system}
\end{figure*}

%メモ: (a)-(c)として界面のスナップショットを追加．密度分布は(d)
\begin{figure*}[htbp]
 \centering
\includegraphics[width=0.8\linewidth]{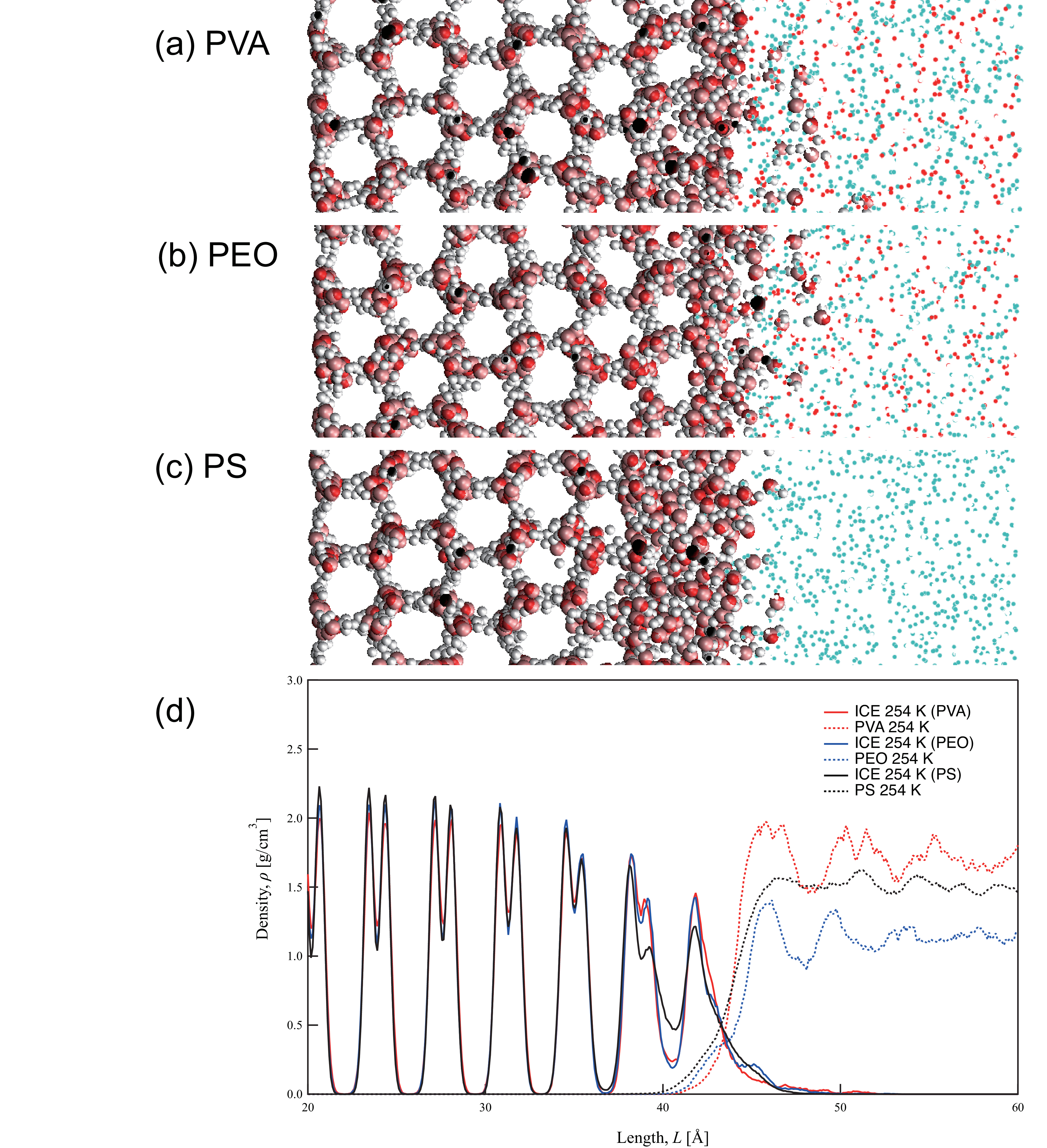}
\caption{Snapshots (a-c) of the polymer-ice interface and density distributions (d) at low temperature (254 K). (a) PVA, (b) PEO, and (c) PS.}
\label{temp_low}
\end{figure*}

%メモ: (a)-(c)として界面のスナップショットを追加．密度分布は(d)
\begin{figure*}[htbp]
 \centering
\includegraphics[width=0.8\linewidth]{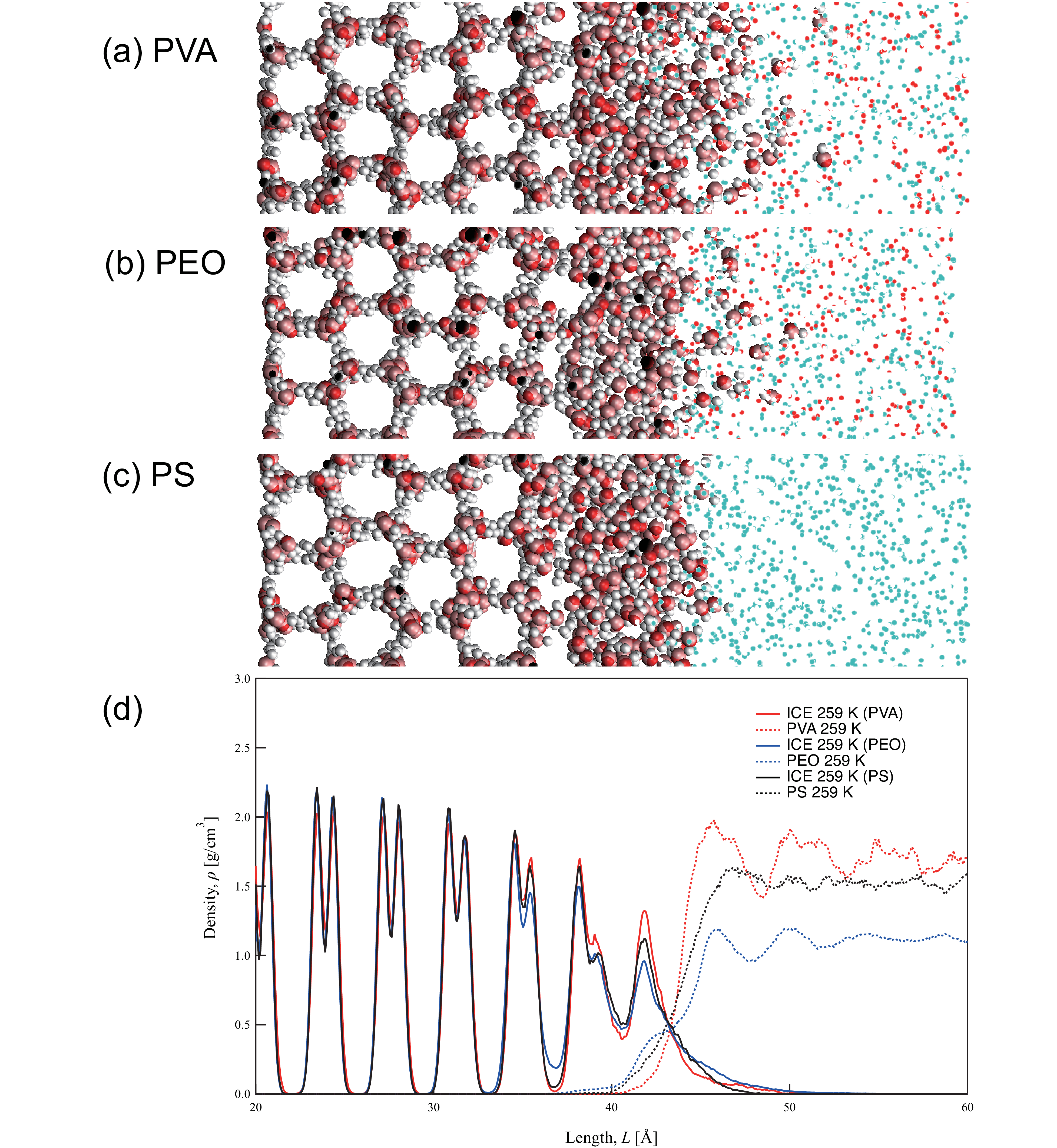}
\caption{Snapshots (a-c) of the polymer-ice interface and density distributions (d) at mid temperature (259 K). (a) PVA, (b) PEO, and (c) PS.}
\label{temp_mid}
\end{figure*}

%メモ: (a)-(c)として界面のスナップショットを追加．密度分布は(d)
\begin{figure*}[htbp]
 \centering
\includegraphics[width=0.8\linewidth]{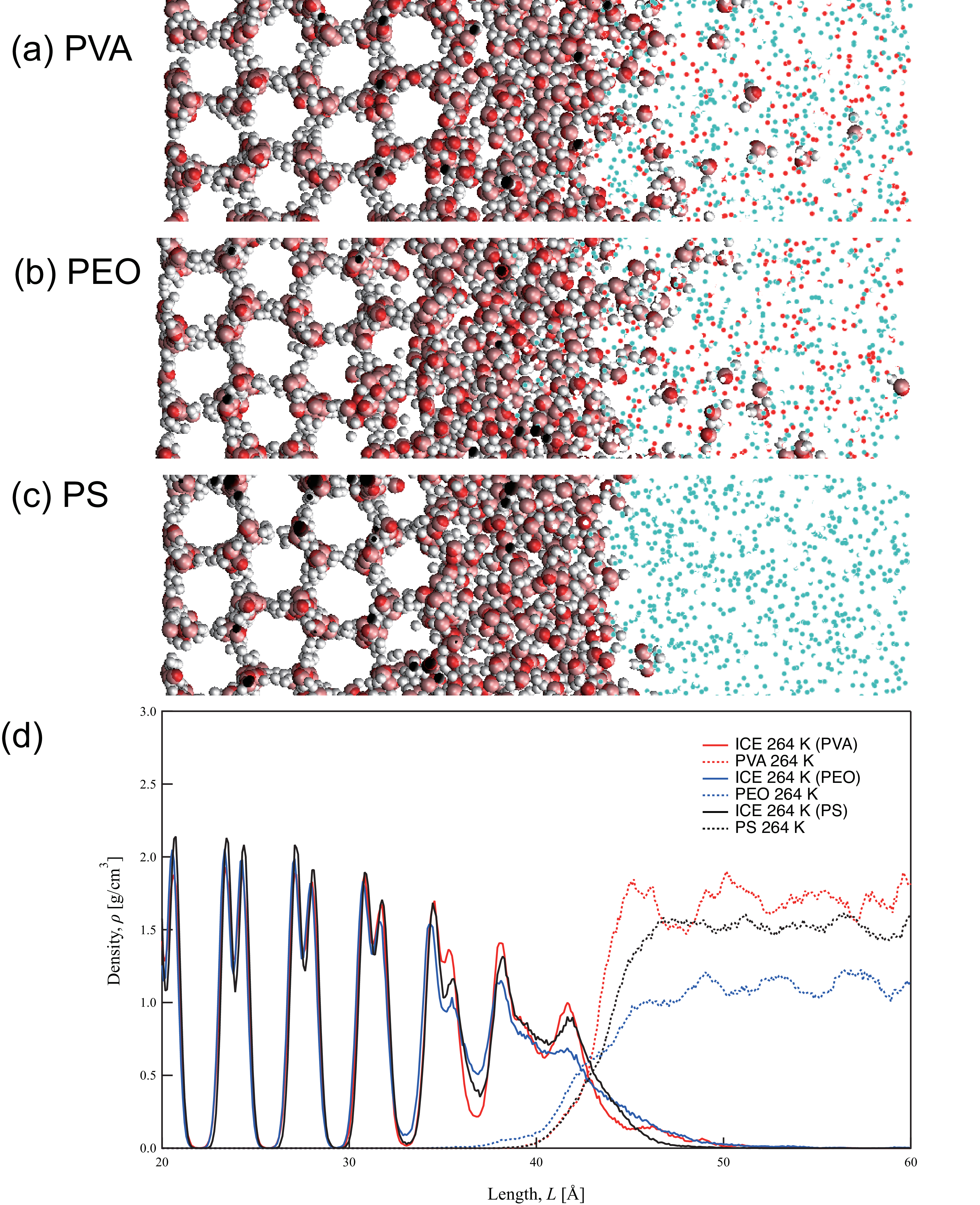}
\caption{Snapshots (a-c) of the polymer-ice interface and density distributions (d) at high temperature (264 K). (a) PVA, (b) PEO, and (c) PS.}
\label{temp_high}
\end{figure*}

\begin{figure*}[htbp]
 \centering
\includegraphics[width=0.8\linewidth]{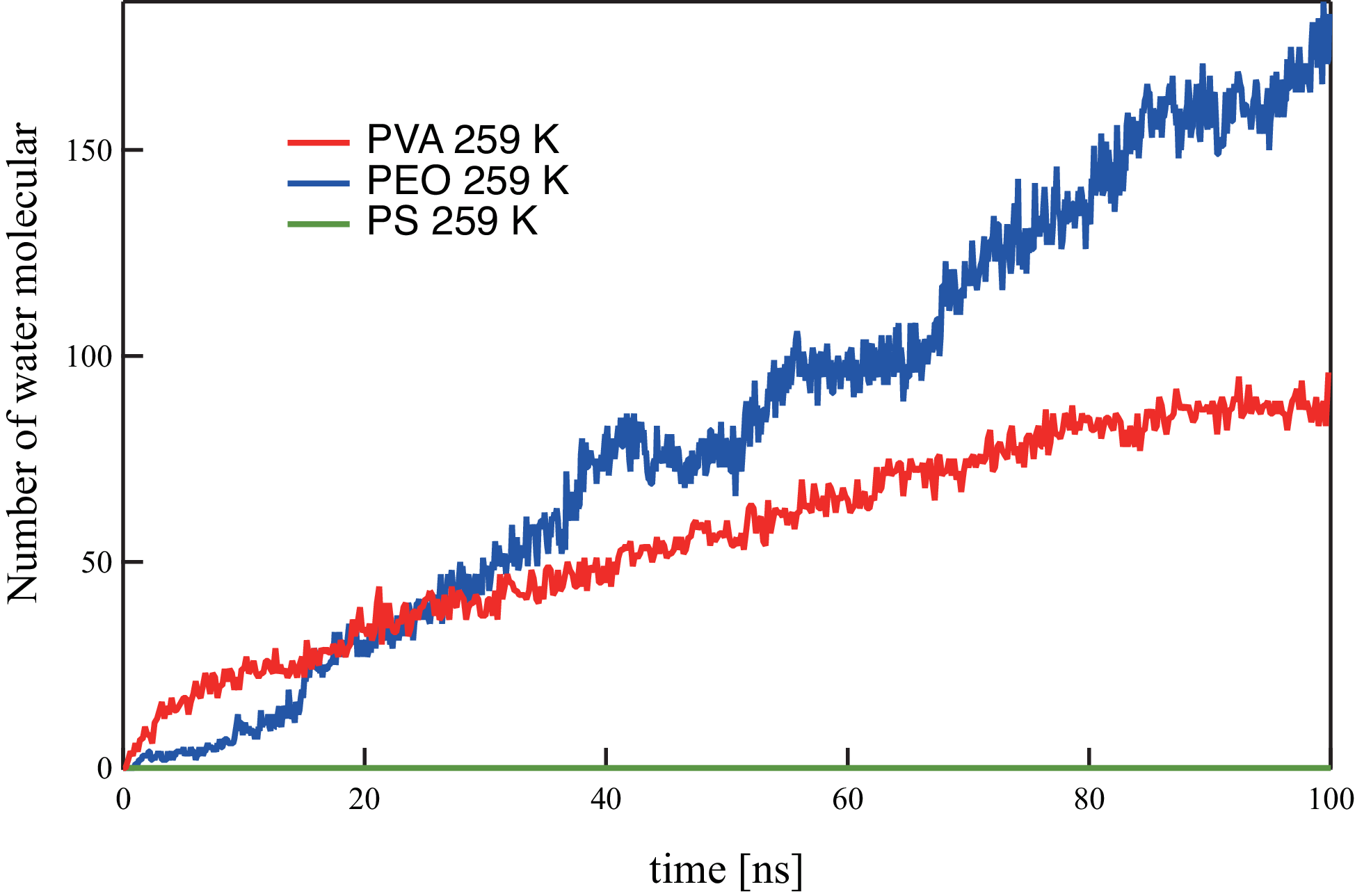}
\caption{Time variation of the number of water molecules entering the polymer side. Using the density distribution of each polymer in Fig. \ref{temp_mid}, the first minimum on the interface side was defined as the boundary, and the water molecules that entered the polymer side were counted.}
\label{num_water}
\end{figure*}

%メモ: 何故か振動している
\begin{figure*}[htbp]
 \centering
\includegraphics[width=0.8\linewidth]{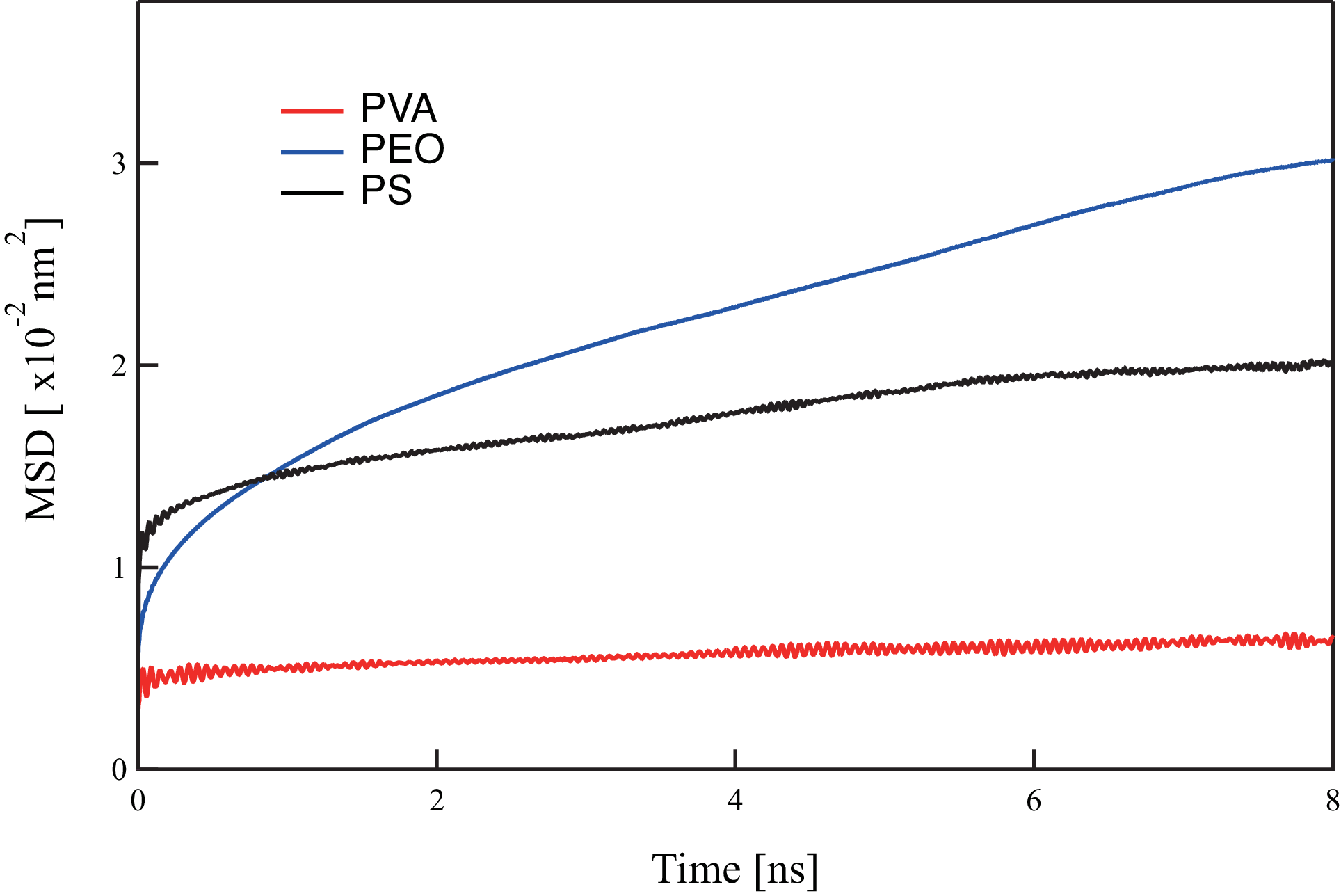}
\caption{Mean square displacement (MSD) for all atoms constituting each polymer.}
\label{msd}
\end{figure*}

\begin{figure*}[htbp]
 \centering
\includegraphics[width=0.8\linewidth]{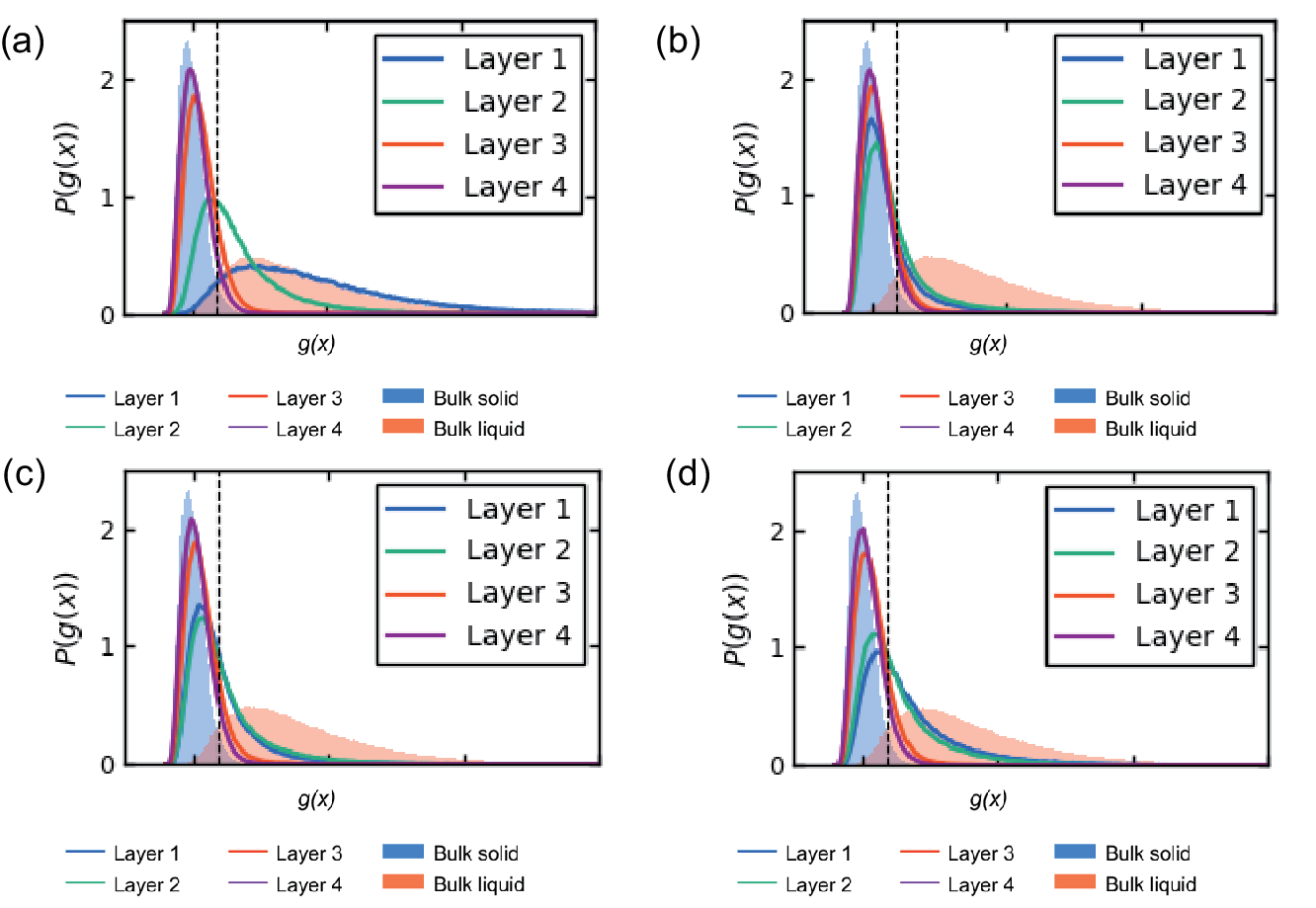}
\caption{Distribution of the liquid-like score, $g(\bm{x})$, for the ice premelting layers at the interfaces with (a) vapor, (b) PVA, (c) PEO, and (d) PS. For comparison, the $g(\bm{x})$ distributions for bulk ice and bulk liquid water are also shown.}
\label{ML}
\end{figure*}

\end{document}